\documentclass[a4paper,useAMS,usenatbib]{mnras}
\usepackage{graphics,epsfig,psfig}
\usepackage{todonotes}
\usepackage[normalem]{ulem}
\usepackage{xcolor}
\usepackage{float}
\usepackage[style=base, singlelinecheck=off, font=small, labelfont=bf, 
justification=raggedright]{caption}
\usepackage{subfigure}
\usepackage{graphicx}
\usepackage[]{inputenc,amssymb}
\usepackage{natbib}
\usepackage{url}
\usepackage{widetext}
\usepackage{amsmath}
\usepackage{cancel}

  \makeatletter
\def\@fnsymbol#1{\ensuremath{\ifcase#1\or \dagger\or \ddagger\or
   \mathsection\or \mathparagraph\or \|\or **\or \dagger\dagger
   \or \ddagger\ddagger \else\@ctrerr\fi}}
    \makeatother
\def \be{\begin{equation}}
\def \ee{\end{equation}}
\def \bea{\begin{eqnarray}}
\def \eea{\end{eqnarray}}

\def\apj{ApJ}
\def\mnras{MNRAS}
\def\prd{PRD}
\def\aap{A\&A}
\definecolor{webgreen}{rgb}{0,.5,0}
\definecolor{webbrown}{rgb}{.6,0,0}

\def\aap{A\&A}
\def\apj{ApJ}
\def\apjs{ApJS}
\def\apjl{ApJL}
\def\mnras{MNRAS}

\def\aj{AJ}
\def\nat{Nature}

\def\prd{Phys.Rev.D}         % Physical Review D
\def\jcap{JCAP}

\title[Delay time \& lensing event rate]{Impact of astrophysical binary coalescence timescales on the rate of lensed gravitational wave events}
\author[Mukherjee, Broadhurst, Diego, Silk, Smoot (2021)]
{Suvodip Mukherjee$^{1,2,3}$\thanks{s.mukherjee@uva.nl}, Tom  Broadhurst$^{4,5,6}$\thanks{tom.j.broadhurst@gmail.com}, Jose M. Diego$^{7}$\thanks{jdiego@ifca.unican.es}, Joseph Silk$^{8,9, 10}$\thanks{silk@iap.fr}, 
\newauthor
\& George F. Smoot$^{5,11,12, 13}$\thanks{gfsmoot@lbl.gov}\thanks{Author list is in the alphabetical order except for the corresponding author.}\\
$^{1}$ Gravitation Astroparticle Physics Amsterdam (GRAPPA),
Anton Pannekoek Institute for Astronomy and Institute for Physics,\\
University of Amsterdam, Science Park 904, 1090 GL Amsterdam, The Netherlands\\
$^2$ Institute Lorentz, Leiden University, PO Box 9506, Leiden 2300 RA, The Netherlands\\
$^3$Delta Institute for Theoretical Physics, Science Park 904, 1090 GL Amsterdam, The Netherlands\\
$^{4}$ Department of Theoretical Physics, University of the Basque Country UPV-EHU, 48040 Bilbao, Spain\\
$^5$ Donostia International Physics Center (DIPC), 20018 Donostia, The Basque Country\\
$^6$ IKERBASQUE, Basque Foundation for Science, Alameda Urquijo, $36-5$ 48008 Bilbao, Spain\\
$^7$ Instituto de F\'isica de Cantabria, CSIC-Universidad de Cantabria, E-39005 Santander, Spain\\
$^{8}$Institut d'Astrophysique de Paris (IAP), UMR 7095, CNRS/UPMC Universit\'e Paris 6, Sorbonne Universit\'es,\\ 98 bis boulevard Arago, F-75014 Paris, France\\
$^9$ The Johns Hopkins University, Department of Physics \& Astronomy, 3400 N. Charles Street, Baltimore, MD 21218, USA\\
$^{10}$ Beecroft Institute for Cosmology and Particle Astrophysics, University of Oxford, Keble Road, Oxford OX1 3RH, UK\\
$^{11}$ IAS TT \& WF Chao Foundation Professor, IAS, Hong Kong University of Science and Technology,\\
Clear Water Bay, Kowloon, 999077 Hong Kong, China\\
$^{12}$ Paris Centre for Cosmological Physics, Universit\'e de Paris, {\it emeritus},
CNRS, Astroparticule et Cosmologie, $F-75013$ Paris, France A,\\
10 rue Alice Domon et Leonie Duquet, 75205 Paris CEDEX 13, France\\
$^{13}$  Physics Department and Lawrence Berkeley National Laboratory, {\it emeritus}, University of California, Berkeley, 94720 CA, USA\\
}
\pubyear{2021}
\begin{document}
\label{firstpage}
\pagerange{\pageref{firstpage}--\pageref{lastpage}}
\maketitle

\label{firstpage}

\begin{abstract}
The expected event rate of lensed gravitational wave sources scales with the merger rate at redshift $z\geq 1$, where the optical depth for lensing is high. It is commonly assumed that the merger rate of the astrophysical compact objects is closely connected with the star formation rate, which peaks around redshift $z\sim 2$. However, a major source of uncertainty is the delay time between the formation and merger of compact objects. We explore the impact of delay time on the lensing event rate. We show that as the delay time increases, the peak of the merger rate of gravitational wave sources gets deferred to a lower redshift.  This leads to a reduction in the event rate of the lensed events which are detectable by the gravitational wave detectors. We show that for a delay time of around $10$ Gyr or larger, the lensed event rate can be less than one per year for the design sensitivity of LIGO/Virgo. We also estimate the merger rate for lensed sub-threshold for different delay time scenarios, finding that for larger delay times the number of lensed sub-threshold events is reduced, whereas for small-delay time models they are significantly more frequent. This analysis shows for the first time that lensing is a complementary probe to explore different formation channels of binary systems by exploiting the lensing event rate from the well-detected events and sub-threshold events which are measurable using the network of gravitational wave detectors. 
\end{abstract}


\section{Introduction}
Gravitational lensing of gravitational waves (GW) is an inevitable consequence due to the intervening matter distribution between the GW source and observer \citep{1992grle.book.....S, Bartelmann:2010fz, PhysRevLett.80.1138, Wang:1996as,Takahashi:2003ix, Dai:2016igl, Broadhurst:2018saj, Diego:2018fzr, Mukherjee:2019wcg, Oguri:2019fix,Mukherjee:2019wfw,Mukherjee:2020tvr}. The lensing of the GWs can lead to the magnification of the strain of the GW signal which can be described by strong or weak lensing depending on strength of the magnification factor. Though there is no confirmed detection of lensed GW sources from O1+O2 \citep{2019ApJ...874L...2H} and O3a observations of the LIGO/Virgo collaboration \citep{Abbott:2021iab}, detection of lensed systems is likely in the future \citep{Dai:2016igl, Ng:2017yiu, Broadhurst:2018saj, Broadhurst:2019ijv, Diego:2019rzc, Mukherjee:2020tvr, Broadhurst:2020moy}. In the future, GW detectors such as Einstein Telescope and Cosmic Explorer will be able to detect a large number of lensed events \citep{2013JCAP...10..022P, 2014JCAP...10..080B, 2015JCAP...12..006D}.

The number of lensed GW events with magnification factor $\mu$ which are detectable is going to depend on the instrument sensitivity, lensing optical depth, and the merger rate of GW sources at high redshift. While the estimation of the instrument sensitivity and lensing optical depth is possible, the largest uncertainty is associated with the merger rate of the GW sources at high redshift. An upper bound on the lensing event rate is possible to impose from the detection (or in the absence of detection) of the amplitude of the stochastic gravitational-wave background as shown by \citet{Mukherjee:2020tvr}. Using the data from O1+O2, \citet{Buscicchio:2020cij, Buscicchio:2020bdq} imposes an upper bound on the lensing event rate for binary black holes and binary neutron star events. 

One of the ways forward to understand the lensing event rate is to motivate the expected event rate for the astrophysical black holes (ABHs) by the star formation rate. However, one of the key aspects to relating the time of the formation of the stars to the time of the black hole merger is the unknown delay time between formation and merger of the black holes \citep{2010ApJ...716..615O,2010MNRAS.402..371B, 2012ApJ...759...52D, Dominik:2014yma, 2016MNRAS.458.2634M, Lamberts:2016txh, 2018MNRAS.474.4997C, Elbert:2017sbr, Eldridge:2018nop, Vitale:2018yhm, Buisson:2020hoq,Santoliquido:2020axb}. In this paper, we investigate the impact of different delay time models between the formation and merger of the ABHs, and their impact on the lensing event rate. For different models of the astrophysical delay time between the formation and merger, we show how the expected lensing rate for different magnifications can be affected. This analysis provides us a more realistic understanding of the expected event rate of the lensed systems.

{By using the strongly lensed events we can infer the high redshift merger rate of the GW sources using both well-detected events as well as using the sub-threshold events. As a result, with the detection of lensed GW events (or even in the absence of detection of lensed GW events), it is possible to infer (or constrain) different delay time scenarios using the lensing statistics. Measurement of different delay times is a direct probe of different formation scenarios of the binary systems. We show in this work for the first time that lensing is a complementary probe to infer the delay time distribution and explore different formation channels of binary systems. This is particularly informative for learning about the high redshift merger rates, which are not accessible from the unlensed low-redshift well-detected events.}

\begin{figure*}
    \centering
    \includegraphics[width=1.\linewidth]{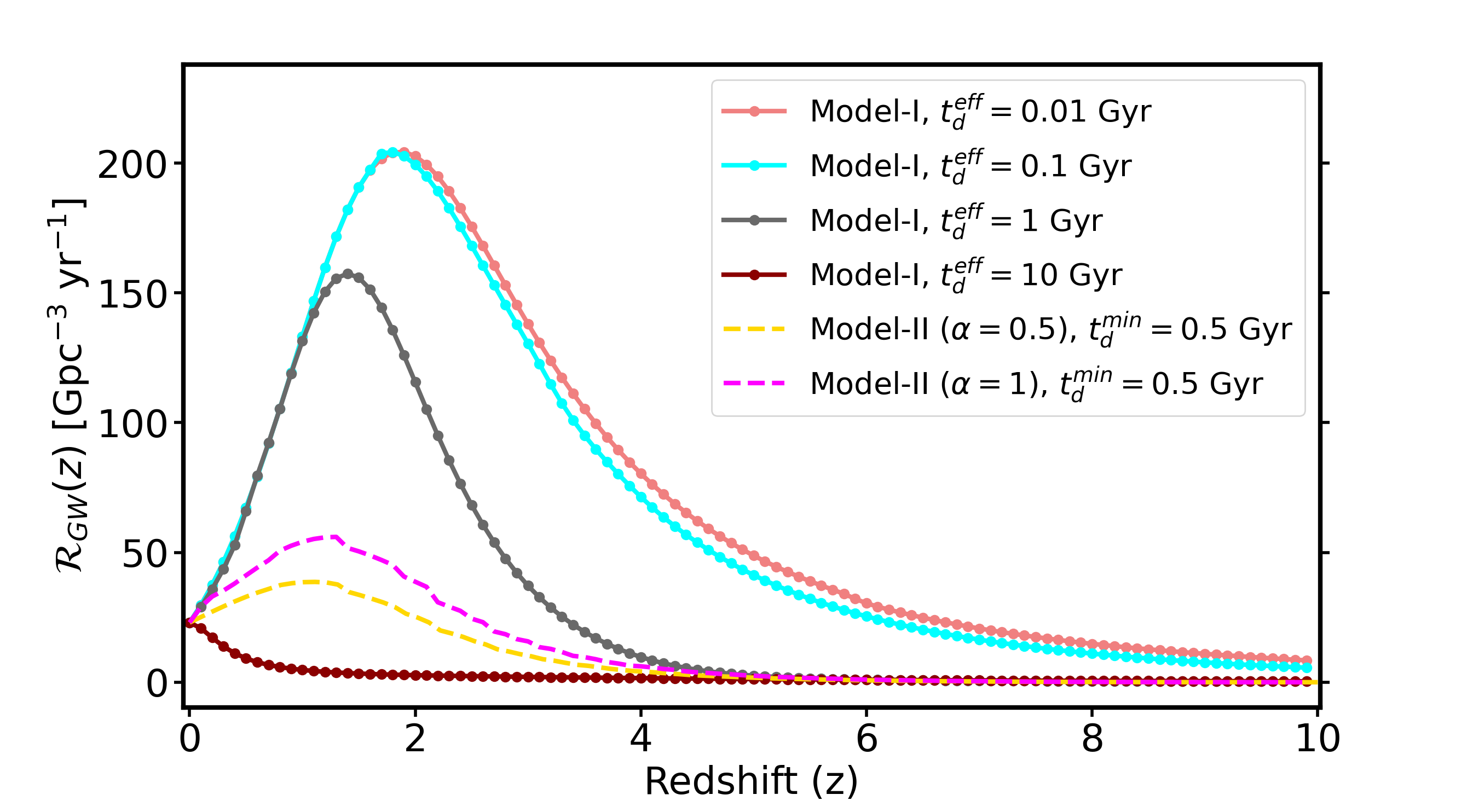}
    \caption{We show the merger rate of the GW sources for different astrophysical delay time models. For sources with small delay time values, the merger rate is higher at a higher redshift, whereas for larger delay time the merger rate at high redshift is significantly reduced. We have chosen the local merger rate of the GW sources according to the estimation from GWTC-2.}
    \label{fig:merger_rate}
\end{figure*}

\begin{figure*}
    \centering
    \includegraphics[width=1.\linewidth]{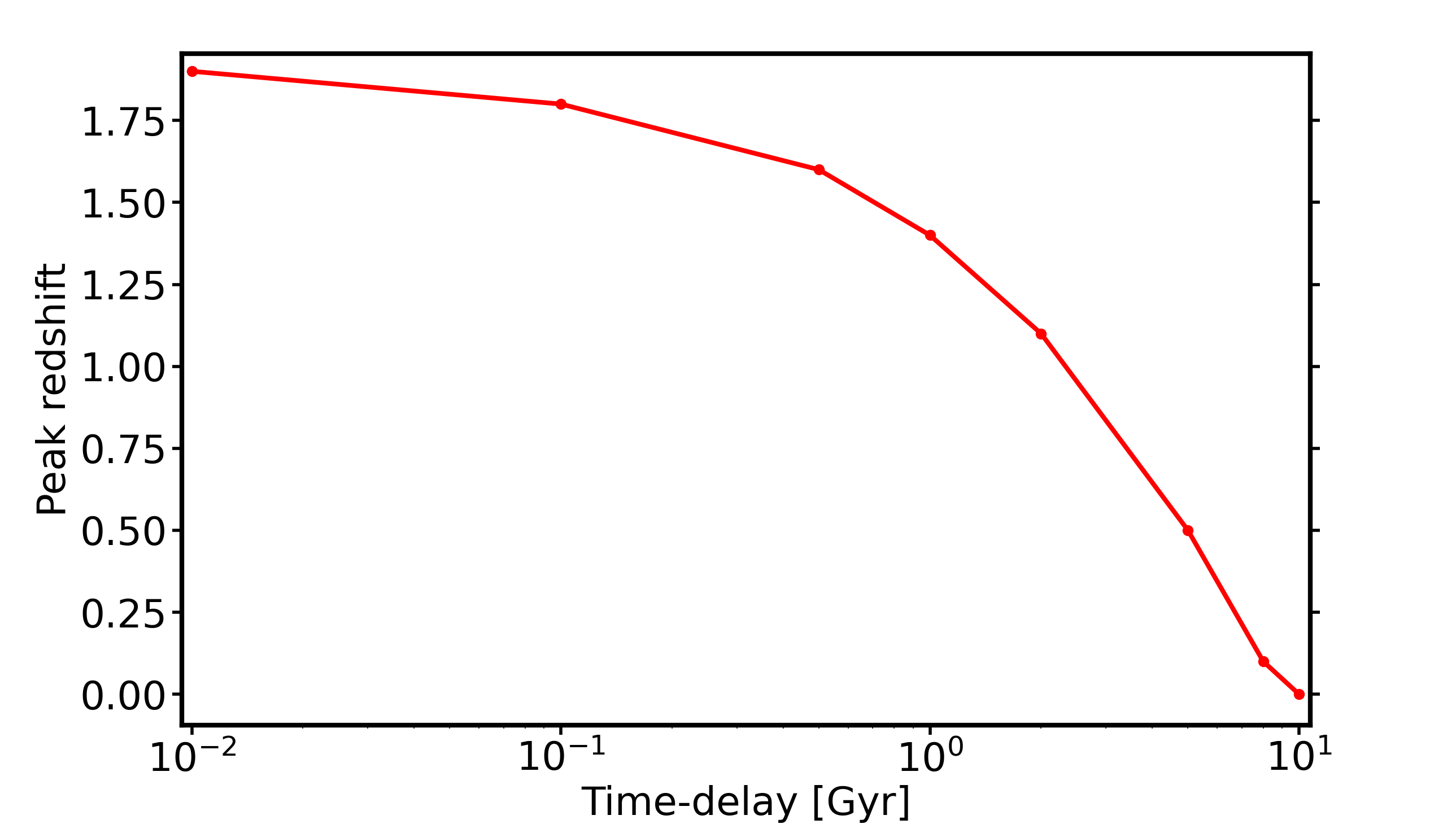}
    \caption{We show the position of the peak redshift of the GW merger rate for different delay time values for the Gaussian model (Model-I).}
    \label{fig:zpeak}
\end{figure*}

\section{Astrophysical delay time between the formation and merger of BBHs}\label{ABHtimedelay}
The probability distribution of the delay time between the formation of the progenitor star and the merger of the black holes is not well known. From population synthesis, it is expected that the delay time can range between a few 100 Myr to about the age of the Universe. The probability distribution on the delay time is not trivially derived and depends on the stellar metallicity $Z_\odot$ of the host galaxy and also on the mass of the BBHs. However, if we are interested in the lensing event rate, marginalized over the mass distribution and stellar metallicity of the host galaxy, then we can define a probability distribution of an effective delay time parameter $t^{eff}_d$ as
\begin{equation}\label{ptd}
    P(t^{eff}_d) = \int d\mathcal M dZ_{\odot} f_{eff}  (\mathcal M, Z_{\odot}) P(t_d (\mathcal M, Z_{\odot})),
\end{equation}
where, $f_{eff} (\mathcal M, Z_\odot)$ is the efficiency with which black holes form and merge of chrip mass $\mathcal{M}$. The probability distribution of the effective-delay time parameter is not yet known from observations and needs to be modeled. 

The total merger rate of GW sources of astrophysical origin  can then be modelled as 
\begin{equation}\label{rate1}
    \mathcal{R}_{GW}(z_m)= \mathcal{N}\int^{\infty}_{z_m} dz \frac{dt_f}{dz}P(t^{eff}_d) R_{SFR}(z),
\end{equation}
where $\mathcal N$ is the normalization such that the merger rate today ($z=0$) agrees with the observations from GWTC-2 $r_{GW}(0)= 23.9^{+14.3}_{-8.6}$ Gpc$^{-3}$ yr$^{-1}$ \citep{Abbott_2020}. $R_{SFR}(z)$ is the star-formation rate which is well described  by the \citep{Madau:2014bja} relation
\begin{equation}\label{mdsfr}
R_{SFR}(z)= 0.015 \frac{(1+z)^{2.7}}{1+ (\frac{(1+z))}{2.9})^{5.6}} \, \text{M}_\odot \text{Mpc}^{-3} \text{yr}^{-1}.
\end{equation}
The model of the merger rate of BBHs in Eq. \eqref{rate1} depends only on the model of the probability distribution for the delay time. The total volume averaged  merger rate of the BBHs can be written as 
\begin{equation}\label{rate2}
    R_{GW}(z)= \int d z_m \frac{ \mathcal{R}_{GW}(z_m)}{1+z_m} \frac{dV}{dz}.
\end{equation}
We consider two models for the delay time probability distribution $ P(t^{eff}_d)$ in our analysis, with the values of the delay time which are motivated by stellar population synthesis namely, 
{\begin{align}\label{ptdop}
\begin{split}
   P(t^{eff}_d)=&  \frac{1}{\sqrt{2\pi}\sigma_{t^{eff}_d}}\exp{\bigg[\frac{-(t_f-t_m-t^{eff}_d)^2}{2\sigma^2_{t^{eff}_d}}\bigg]}\hspace{0.5cm} \text{Model-I},\\
   P(t^{eff}_d)=& (1/{t^{eff}_d})^\alpha, \,\, t^{eff}_d > t^{min}_d, \alpha \in\{0.5, 1\} \hspace{0.6cm} \text{Model-II},\\
\end{split}
\end{align}}
where, $t_f$ is the time of formation of the star and $t_m$ is the time of merger, the value of $\sigma_{t^{eff}_d}$ is considered as $20\%$ of the mean value in this analysis. We consider a minimum delay time of 10 Myr in this paper and a maximum delay time of the age of the Universe, to make sure that the BBHs are going to merge within the Hubble time. The models with a longer delay time lead to a greater shift in the peak of the BBH distribution value to a lower redshift from the SFR peak (which is around redshift $z_p\approx 2$). In contrast, for smaller delay time, the BBH merger rate follows the Madau-Dickinson redshift dependence in the lower redshift and the peak shifts to a lower value than $z_p=2$. We show the variation of the BBH merger rate for different scenarios in Fig. \ref{fig:merger_rate}. {The  delay time distribution between formation and merger depends on the formation channel \citep{2010ApJ...716..615O,2010MNRAS.402..371B, 2012ApJ...759...52D, Dominik:2014yma, 2016MNRAS.458.2634M, Lamberts:2016txh, 2018MNRAS.474.4997C, Elbert:2017sbr, Eldridge:2018nop, Vitale:2018yhm, Buisson:2020hoq,Santoliquido:2020axb}, and its imprint on the GW merger rate for different form of the probability distribution is shown in Fig. \ref{fig:merger_rate}. The scenarios with a delay time distribution $1/t_{d}^{eff}$ is expected for the formation scenarios if the distribution of the initial binary separation is flat in %the 
log-space \citep{2010ApJ...716..615O, 2012ApJ...759...52D}. However, there is a significant variation in the delay time distribution %is 
possible for other distributions of the initial separation, stellar metallicity, and property of the host galaxy \citep{Lamberts:2016txh, 2018MNRAS.474.4997C, Elbert:2017sbr, Eldridge:2018nop, Vitale:2018yhm, Buisson:2020hoq,Santoliquido:2020axb}. The scenario with $1/t_{d}^{eff}$ and a minimum delay time of $0.5$ Gyr is usually assumed as a standard scenario from stellar population synthesis \citep{2010ApJ...716..615O,2010MNRAS.402..371B, 2012ApJ...759...52D}.} In Fig. \ref{fig:zpeak}, we show the peak position of the GW merger rates for the Gaussian model (Model-I) as a function of the mean value of the delay time. For the power-law model (Model-II) with a fixed minimum value of the delay time of $0.5$ Gyr, the GW merger rate peaks around redshift $z\sim 1.1$ for $\alpha=0.5$ and $z\sim 1.2$ for $\alpha=1$ respectively. 

\begin{figure*}
    \centering
    \includegraphics[width=1.\linewidth]{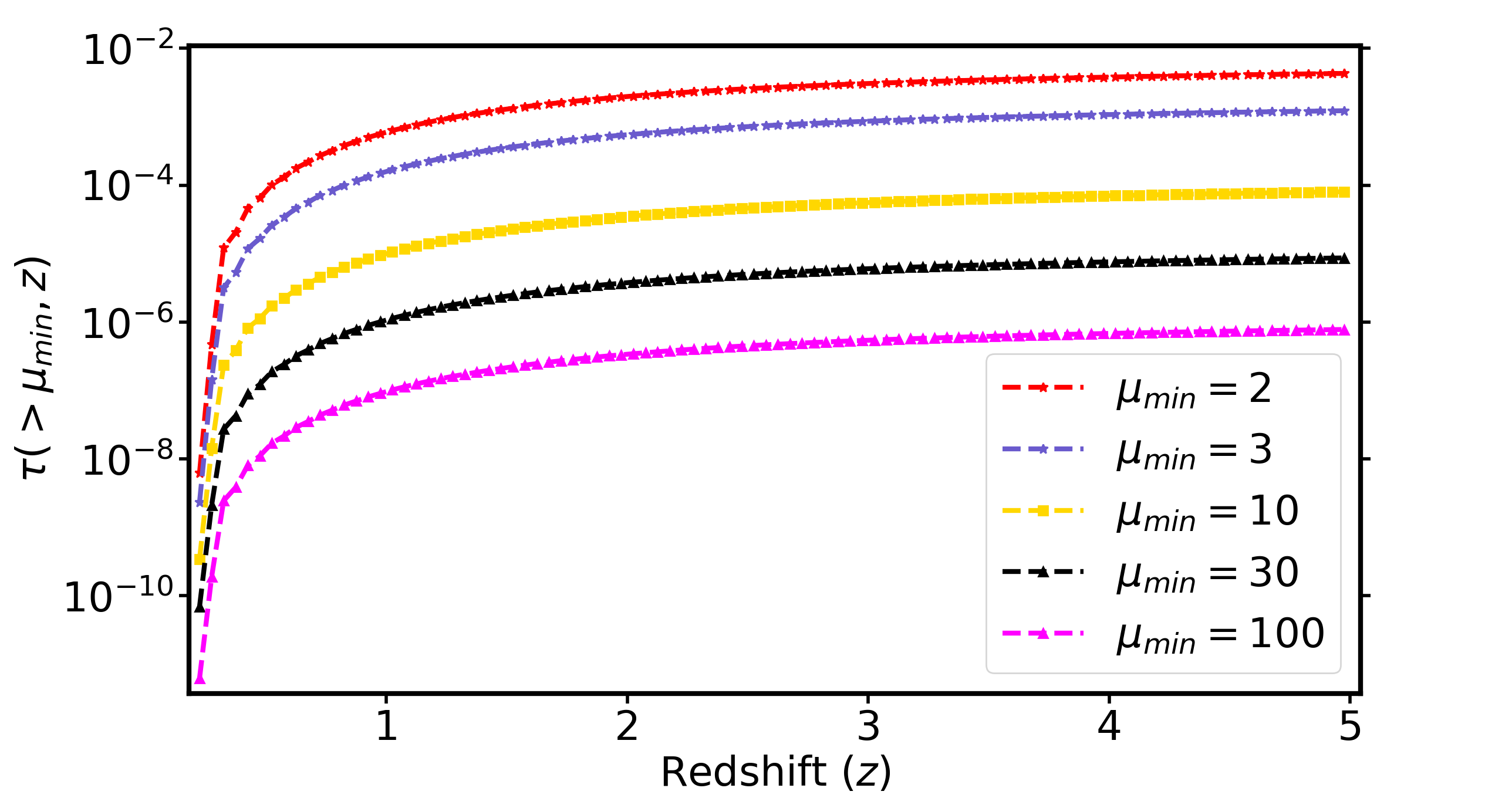}
    \caption{The lensing optical depth $\tau$ as a function of redshift for different magnification factors is shown. The lensing optical depth is an increasing function of redshift and saturates above redshift $z=2$.}
    \label{fig:tauz}
\end{figure*}

\section{Impact of astrophysical delay time on the lensing event rate of GW sources}
Gravitational lensing of GWs due to the intervening structure leads to lensed GW sources. Lensing of GWs leads to magnification of the GW strain which can be written  {in the inspiral phase of the coalescing binaries} as \citep{1987thyg.book.....H, Cutler:1994ys,Poisson:1995ef,maggiore2008gravitational}
\begin{align}\label{lens-1}
    \begin{split}
       h_{\pm}(f) (\hat n)= \sqrt{\mu}\sqrt{\frac{5}{96}}\frac{G^{5/6}\mathcal{M}_z^2 (f_z\mathcal{M}_z)^{-7/6}}{c^{3/2}\pi^{2/3}d_L}\mathcal{I}_{\pm} (\hat L.\hat n), 
           \end{split}
\end{align}
where $\mu$ is the magnification factor, $f_z= f/(1+z)$ is the redshifted frequency, $d_L$ is the luminosity distance to the gravitational wave source, $\mathcal{M}_z= (1+z)\mathcal{M}$ is the redshifted chirp mass, and  $\mathcal{I}_{\pm} (\hat L.\hat n)$ captures the projection of the angular momentum vector $\hat L$ on the line of sight $\hat n$.  {After the inspiral phase of the coalescing binaries, it goes through the merger and ringdown phase, which we model according to the IMRPhenomA model \citep{Ajith:2007kx}.} The observed GW signal  {for all the three phases (inspiral, merger, and ringdown),} after including the detector response function $F_{\pm} (\theta, \phi, \psi)$ can be written as $h_{obs}= \sum_{+,-} F_{\pm (\theta, \phi, \psi)} h_{\pm}$. The detection of a GW signal depends on the matched-filtering signal to noise ratio $\rho$ which can be obtained by taking an inner product with the GW waveform $W(f)$ as \citep{Sathyaprakash:1991mt, Cutler:1994ys,Balasubramanian:1995bm}
\begin{equation}\label{snrgw}
    \rho^2\equiv 4Re\bigg[\int_0^{f_{max}} df \frac{ h^{obs}(f)W^*(f)}{S_n(f)}\bigg],
\end{equation}
where $S_n(f)$ is the noise power spectrum. A GW signal with $\rho>\rho_{th}$ is considered to be a detected signal, so, we can define the detector response function $\mathcal{S}(\theta,z_s, \mu)\equiv H (\rho (\mu, d_L, \mathcal{M}_c)-\rho_{th})$ \footnote{Heaviside step function $H(x)=1$, only when the argument satisfies the criterion $x\geq 1$.} which ensures that only the gravitational wave sources for which $\rho (\mu, d_L, \mathcal{M}_c) \geq \rho_{th}$, can be detected as individual events. \footnote{We have considered the value of $\rho_{th}=8$.}  {In this analysis, we model the strain of the GW signal following \citep{Ajith:2007kx} (which is  known as the IMRPhenomA model})  {and include all the three phases inspiral, merger and ringdown of the waveform \footnote{ {We assume here that the IMRPhenomA waveform can reliably model the inspiral, merger and the ringdown phase for LVK detectors, and we can extract the GW source properties using this waveform. Effect of waveform systematic on the lensing event rate will be explored in a future work.}}. The value of $f_{max}$ in Eq. \eqref{snrgw} is chosen  as $f_{max}= f_{cut}\equiv c^3(a_1\eta^2+a_2\eta+a_3) /(G\pi M)$ in this analysis \citep{Ajith:2007kx},} where $\eta= m_1m_2/M^2$ is the symmetric mass ratio and $M=m_1+m_2$ is the total mass of the GW system.  

The expected number of lensed GW sources which can be detected depends on the lensing optical depth denoted by $\tau$, the merger rate of GW sources $\mathcal{R}_{GW}(z)$, and the detector response function.  $\mathcal{S}(\theta,z_s, \mu)$. It can be estimated as 
\begin{align}\label{nzl-1}
    \begin{split}
     \dot N_l (\geq \mu) \equiv  \frac{dN}{dt} (\geq \mu, z_s)= & \int_0^{z_s}dz \int d\theta  \overbrace{\frac{dV}{dz} \tau_l(\geq \mu, z)}^{\text{Cosmology}}\\ &\times  \overbrace{p(\theta) \frac{\mathcal{R}_{GW}(z)}{(1+z)}}^{\text{Astrophysics}} \overbrace{\mathcal{S }(\theta,z, \mu)}^{\text{Detector response}} ,\\
     %=& \int_0^{z_s} dz\, \mathcal{K}(z, \mu, \theta) \frac{R(z, \theta)}{1+z}
           \end{split}
\end{align}
where, the redshift of the GW source is denoted by $z_s$, differential volume factor at a comoving distance is denoted by $dV/dz$, the probability distribution of the GW source parameters is denoted by $p(\theta)$. The lensing optical depth is defined as  
\begin{align}\label{nzl-2}
    \begin{split}
      \tau_l(\geq \mu, z)= \int_0^z dz_l \frac{d\tau(\geq \mu, z, z_l)}{dz_l},
           \end{split}
\end{align}
where, $d\tau(\geq \mu, z, z_l) / dz_l$ %$\frac{d\tau(\geq \mu, z, z_l)}{dz_l}$ 
is the differential lensing optical depth which can be written as \citep{Turner:1984ch}
\begin{align}\label{nzl-2a}
    \begin{split}
     \frac{d\tau(\geq \mu, z, z_l)}{dz_l}= \frac{1}{A_T(z_s)} \frac{dV(z_l)}{dz_l}\int dM \frac{dN}{dMdz} A_N(\mu, M, z_l, z_s),
           \end{split}
\end{align}
where, $A_T(z_s)$ is the area of the spherical shell at redshift $z_s$ in physical units, $\frac{dV(z_l)}{dz_l}$ is the differential volume fraction at the redshift of the lens $z_l$, $\frac{dN}{dMdz}$ is the halo mass function per unit halo mass and redshift, and $A_N(\mu, M, z_l, z_s)$ is the area for magnification higher than $\mu$ computed in the image plane, but divided by the factor $\mu$ to account for the equivalent area in the source plane. The estimation of the lensing optical depth as a function of the magnification factor and cosmological redshift is shown in Fig. \ref{fig:tauz}. {More details on the estimation of the lensing optical depth can be found in these \citep{Watson:2013mea, Diego:2018fzr, Diego:2019rzc}.} 

Eq. \eqref{nzl-1} indicates that the total number of lensed events detectable depends on the interplay between the lensing optical depth, the merger rate of the BBHs, and the detector response function. The main part which decides the number of the detectable lensed event is the overlap in the redshift distribution of the GW merger rate and the lensing optical depth when the detector response function (i.e. $\mathcal{S}(\theta, z_s, \mu)$) is one.  

\begin{figure*}
\centering
\subfigure[ ]{\label{fig:mu2} 
\includegraphics[trim={0.cm 0.cm 0.cm 0.cm},clip,width=1.\linewidth]{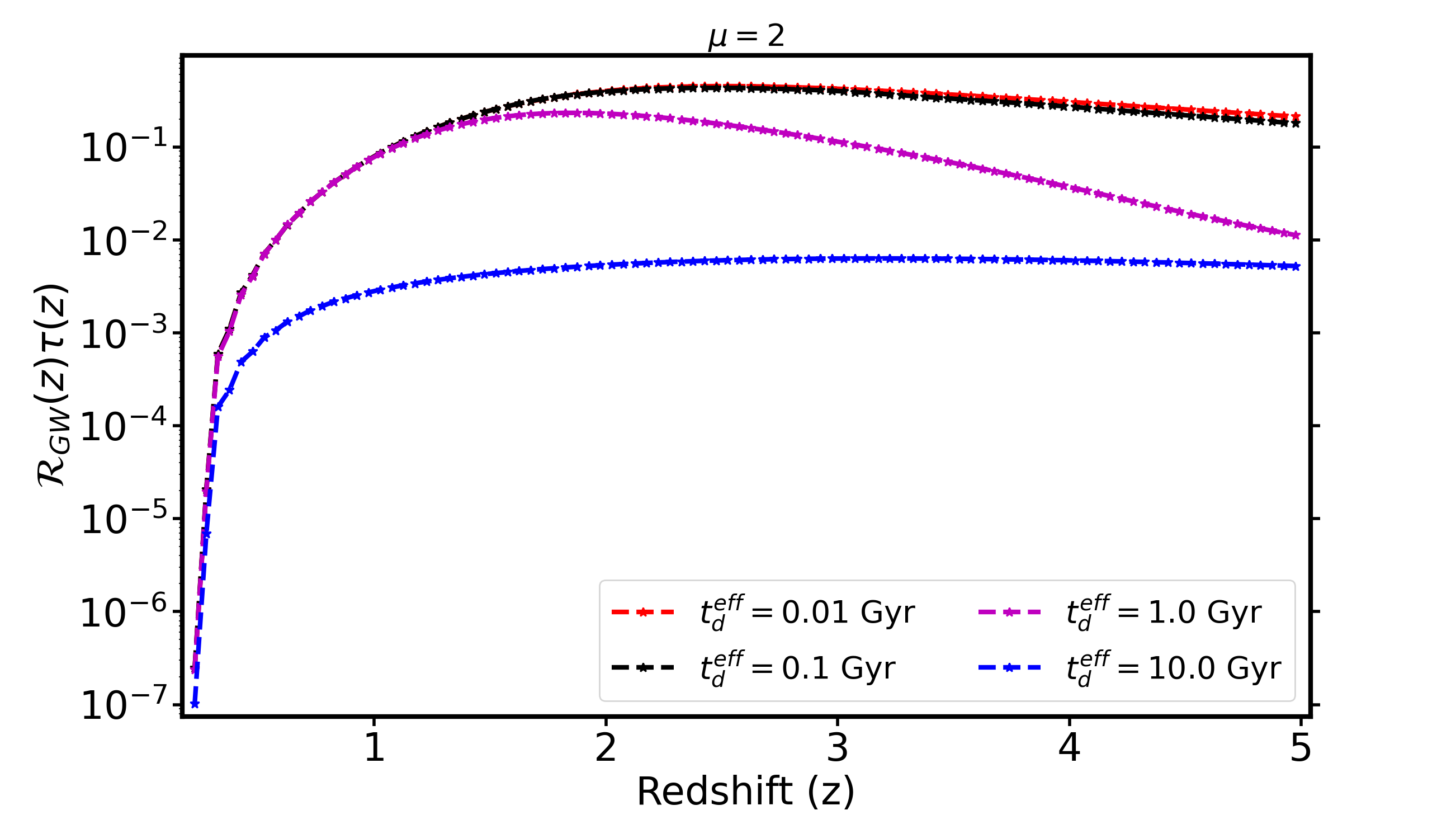}}

\subfigure[ ]{\label{fig:mu30} 
\includegraphics[trim={0.cm 0.cm 0.cm 0.cm},clip,width=1.\linewidth]{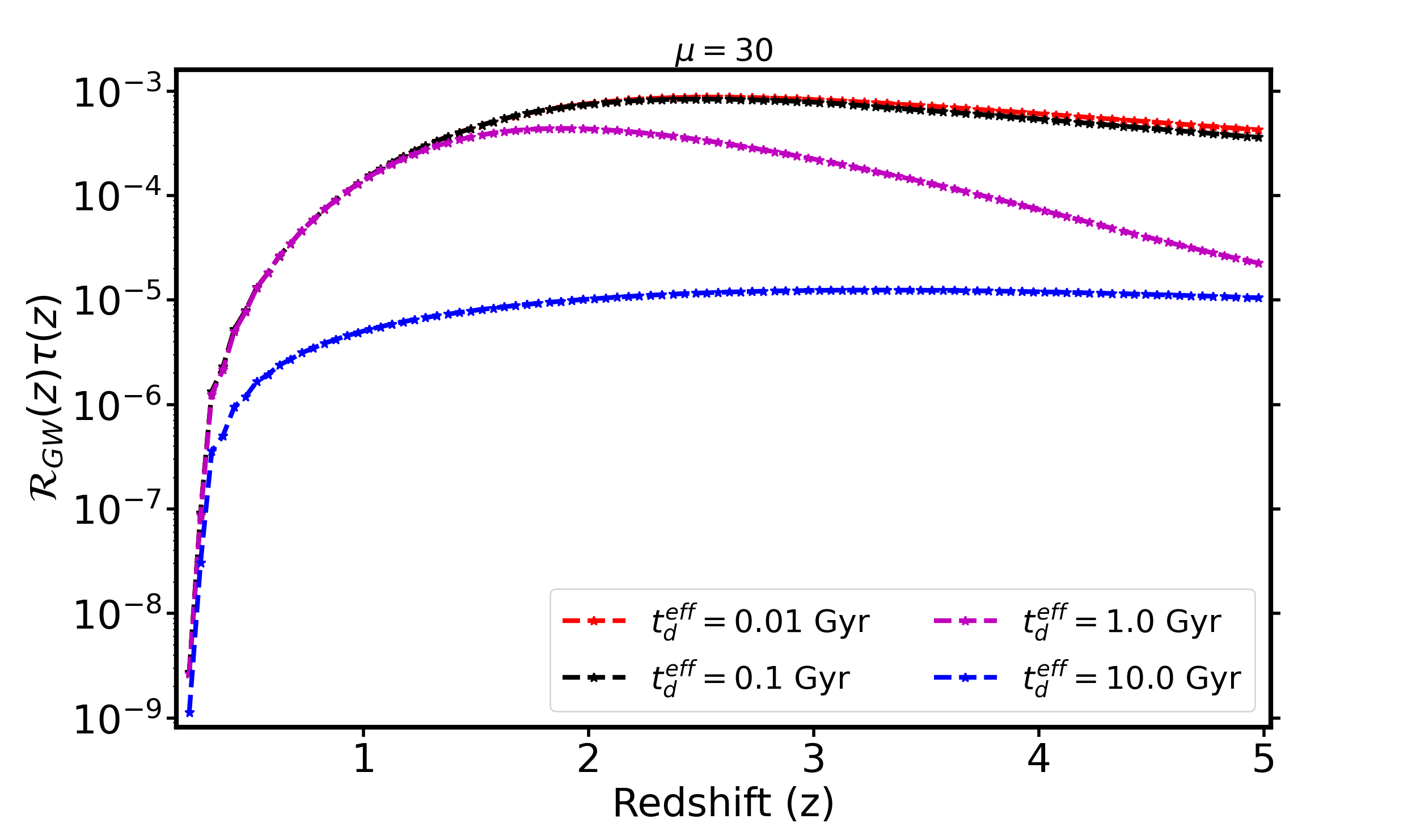}}
\captionsetup{singlelinecheck=on,justification=raggedright}
\caption{We show the product between the GW merger rate for Model-I and the lensing optical depth for the magnification factor (a) $\mu=2$ and (b) $\mu=30$ for different values of the delay time parameter.}
\label{fig:m2}
\end{figure*}

In the presence of delay time between the formation of a star and the merger of BBHs, the redshift distribution of the merger shifts towards a low redshift. For a delay time greater than 1 Gyr, the merger rate decreases for redshift $z\sim 1$. Whereas, the lensing optical depth is an increasing function of redshift (as shown in Fig. \ref{fig:tauz}).  {So, the product between the lensing optical depth and merger rate reduces if the delay time between the formation and the merger is large, whereas when the delay time between formation and merger is small the product between them increases. We have shown the product between the merger rate and the lensing optical depth in Fig. \ref{fig:mu2} and Fig. \ref{fig:mu30} respectively for $\mu=2$ and $\mu=30$, as two representative cases. The value of magnification factor $\mu=2$ is representative of lower redshift events (typically $z=1$ or less), while $\mu=30$ is more representative of higher redshift events ($z=2$ or larger). The nature of the plot will remain the same also for the higher magnification factor. The only change for higher magnification factors will be reflected in the overall amplitude of the product of the signal, which will be governed by the value of $\tau$.  If the product between the lensing optical depth and GW merger rate is less (or more), then the total number of detectable lensing events reduces (or increases).} We examine in the next section the event rate of lensed GW sources for the LIGO-design sensitivity \footnote{ \href{https://dcc.ligo.org/LIGO-T1800044/public}{https://dcc.ligo.org/LIGO-T1800044/public}} \citep{Aasi:2013wya, TheLIGOScientific:2014jea,TheVirgo:2014hva}.

\begin{figure*}
    \centering
    \includegraphics[width=1.\linewidth]{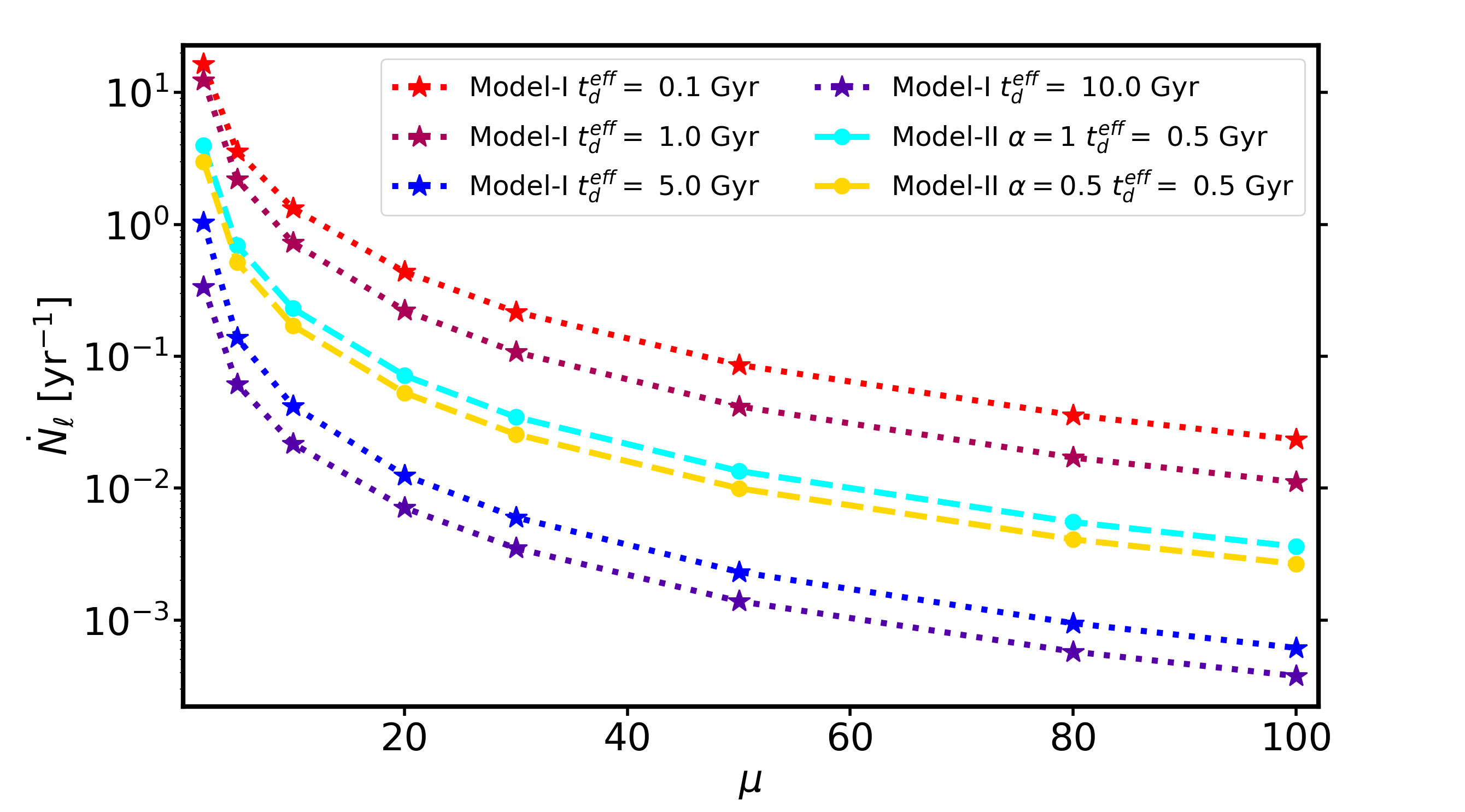}
    \caption{We plot the expected lensing event rate as a function of magnification for different values of the delay time parameters, which can be detected with the matched-filtering SNR $\rho_{th} \geq 8$ with the LIGO-design sensitivity.}
    \label{fig:rate}
\end{figure*}

\section{Lensing event rate for LIGO-design sensitivity }\label{lensingrate}
We calculate the total number of lensed GW events for LIGO-design sensitivity \citep{Aasi:2013wya, TheLIGOScientific:2014jea}. We consider simulated GW signals using the analytical GW waveform from \citep{Ajith:2007kx} for sources in the mass range $m_i \in \{5,50\}$ M$_\odot$ with a power-law distribution on both the masses $m_i \propto 1/m^{2.35}_i$. The GW sources are considered to be non-spinning and the inclination angle is random. Using the probability distribution of the effective angle parameter $\Theta= 2 [F^2_+ (1+\cos^2{i})^2 + 4F^2_\times\cos^2{i}]^{1/2}$ \citep{Finn:1992xs, Finn:1995ah} as $P(\Theta)= 5\Theta(4-\Theta)^3/256$ for $\Theta \in [0,4]$ \citep{Finn:1992xs, Finn:1995ah}, we integrate over the distribution of $\Theta$ to estimate the number of lensed events. The corresponding plot as a function of the lensing magnification factor for different models of delay time is shown in Fig. \ref{fig:rate} for both the models. For models with a small delay time, the number of detectable lensed events is large in comparison to the models with a larger delay time. For the models with variation in the delay time, the event rate of lensing can vary by order of magnitude even for the same detector configuration. This variation in the lensing event rate is directly related to the high redshift merger rate (see Fig. \ref{fig:merger_rate}).

The estimation of the lensing event rate due to delay time is modeled in this analysis as an effective parameter after integrating the mass dependence and the metallicity dependence. However, this is going to be an additional variation in the delay time parameter depending on the GW source properties and the environment of its formation. 

\begin{figure*}
\centering
%\subfigure[ ]{\label{fig:mu2} 
\includegraphics[trim={0.cm 0.cm 0.cm 0.cm},clip,width=1.\linewidth]{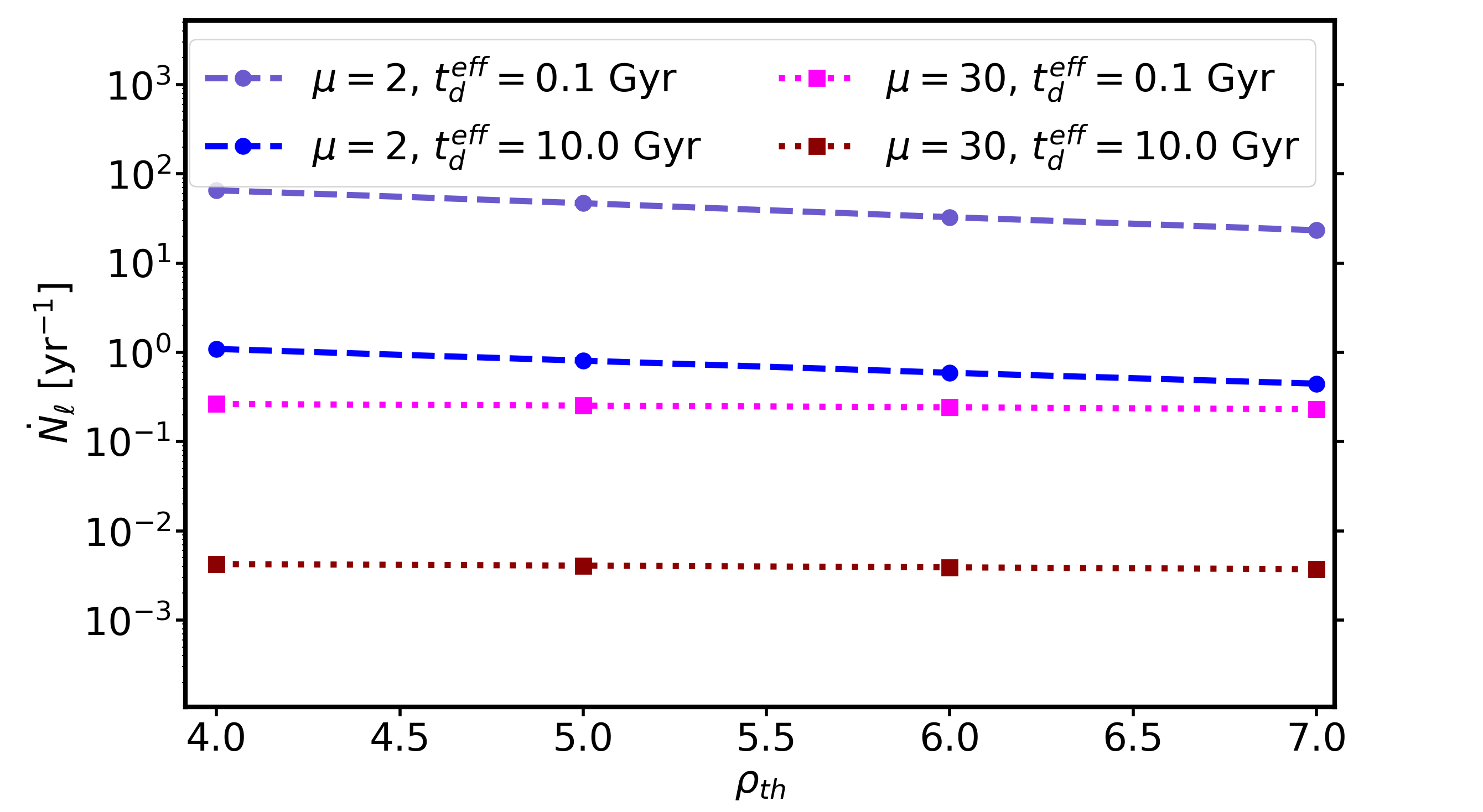}

%\subfigure[ ]{\label{fig:mu30} 
%\includegraphics[trim={0.cm 0.cm 0.cm 0.cm},clip,width=1.\linewidth]{lensing_rate_sub_mu_30.png}}
%\captionsetup{singlelinecheck=on,justification=raggedright}
\caption{We show the lensing event rate as a function of matched-filtering SNR $\rho_{th}$,  for different values of the magnification factor $\mu=2$ and  $\mu=30$ and for different values of the delay time parameter $t^{eff}_d= 0.1$ Gyr and $t^{eff}_d= 10$ Gyr.}
\label{fig:sub_th2}
\end{figure*}

\section{Lensing event rate for sub-threshold events at LIGO-design sensitivity }\label{lensingrate_sub}

The shift in the peak of the GW merger rate to a lower redshift leads to a reduction in the lensed event rate and hence one may need to wait longer to detect a lensed GW source. However, one of the promising avenues to search for the lensed events is to look for the sub-threshold lensed signals. The sub-threshold lensed signal is the lensing events for which the matched-filtering signal to noise ratio of the detection of the signal is $4\leq\rho<8$. For the sub-threshold events with a lower detection threshold and the same magnification, lensed events from high redshifts can be detected.  {The change in the number of lensed GW events with change in the detection threshold depends on the product of the GW merger rate $ R_{GW}(z)$ and the lensing optical depth, which are shown in Fig. \ref{fig:sub_th2} for $\mu=2$ and $\mu=30$ for Model-I. We have chosen the maximum magnification factor $\mu=30$ as a representative value for showing the variation of the signal with different detection thresholds. For $\mu=30$, the sub-threshold detection is $\mathcal{O}(10^0)$ per year, for the small values of the time-delay parameter. For the higher magnification factors, the sub-threshold lensing rates will be even smaller.} So, by using the sub-threshold lensing event rate, we can directly probe the redshift evolution of the  GW merger rate and hence the formation channel of the binary systems. 

Using Eq. \eqref{nzl-1}, we estimate the number of expected lensed events for different sub-threshold cutoffs $\rho_{th}$.  In Fig. \ref{fig:sub_th2}, we show the expected lensing event rate for the sub-threshold events for different delay time models for the magnification factors $\mu=2$ and $\mu=30$ respectively. For a fixed value of the magnification factor $\mu$, if the GW  merger rate increases with redshift, then the number of sub-threshold lensed events also increases with redshift. Whereas if the number of lensing events decreases at high redshift (like for the case with delay time $t_d= 10$ Gyr), then the event rate of sub-threshold lensed sources is also less, even for the same magnification factor, as shown in Fig. \ref{fig:sub_th2}. As a result, the population of the sub-threshold lensing events for different magnification factors can provide direct access to the high redshift merger of the GW sources. In a future analysis, we will show how one can use sub-threshold events to reconstruct the merger of GW sources at higher redshift. {Several data analysis methods are developed to search for sub-threshold lensed events from the GW data \citep{Li:2019osa, McIsaac:2019use}. By applying these techniques to the future data of LIGO/Virgo, we can search for the sub-threshold lensed events and can explore the high redshift of the merger rate of GW sources.}  

\section{Conclusions}

We have shown that predictions for the number of lensed gravitational wave events are sensitive to the delay time between binary formation and coalescence controlling the redshift evolution of detectable events.  If the delay time between formation and merger of the GW sources is large, then the peak merger rate shifts to lower redshift which in turn means fewer lensed GW events are predicted as magnification by lensing. This accesses higher redshifts, $z>1$, where the optical depth for lensing peaks. Hence, if lensing is to be significant, then a relatively short delay time-scale will be implied for BBH merger events.  

  We have estimated the impact of the delay time on the lensing event rate for the LIGO design sensitivity for different models of delay time ranging from $10$ Myr to $10$ Gyr encompassed by stellar population synthesis \citep{2010ApJ...716..615O,2010MNRAS.402..371B, 2012ApJ...759...52D, Dominik:2014yma, Lamberts:2016txh, Elbert:2017sbr, Eldridge:2018nop,  Buisson:2020hoq, Santoliquido:2020axb}.  We show that the expected event rate of lensed events can vary by order of magnitude depending on the delay time between the formation and merger of compact objects. If the delay time is more than a few Gyrs, then most of the binary mergers happen at redshift below $z\approx1$, and for which the probability of lensing is suppressed by the low optical depth. Whereas, if the delay time is less, then the peak of the merger rate of the GW sources is going to be close to the star formation peak. As a result, the event rate of lensed sources can be high. Along with the well-detected lensed events, there can also be sub-threshold lensed GW sources. We estimate the sub-threshold lensed events detectable from the LIGO-design sensitivity for different delay time models. With the sub-threshold events, we can detect the merger rates from the sources which are at higher redshift. If the GW merger rates increase at high redshift (or for the short delay time scenarios), then the event rate of the lensed sub-threshold events is going to be large. In the opposite limit, when the delay time is large, then the expected sub-threshold lensed events are going to small.

The analysis presented in this paper makes it possible to estimate the lensed events for future GW observations which are motivated by the astrophysical merger rate including the delay time. By measuring the lensed events (or in the absence of a detection), we can infer the high redshift merger rate of the GW sources. By using both well detected and sub-threshold events, we can make a reconstruction of the high redshift merger rate of the GW sources. By combining the number of GW sources with a different magnification factor, we can estimate the typical delay time values. A detailed method for estimating the high redshift merger rate from the population of lensed sources will be studied in future work. 

One of the key aspects which the study of the detected lensed events and the sub-threshold lensed events can bring to us is the information of the merger rate of high redshift sources. As shown in Fig. \ref{fig:rate} and Fig. \ref{fig:sub_th2}, the lensing event rate varies very strongly with the value of delay time. So assuming the formation of BHs traces the star formation rate, and given the local GW merger rate, we can infer the delay time parameter from the observed (and sub-threshold) lensed events,  which can be detected with different magnification factors by the LIGO/Virgo detectors in its design sensitivity \citep{Aasi:2013wya, TheLIGOScientific:2014jea, TheVirgo:2014hva}. Vice versa, if the delay time is known from local observations, and assuming this is maintained over time, one could use the observed rate of lensed events to infer the intrinsic rate of BH formation at redshifts beyond the reach of the detector sensitivity (thanks to magnification). 

Our calculations are in the context of evolution set by the empirical Madau Dickinson relation fitted to the measured evolution of the integrated star formation rate with redshift \citep{Madau:2014bja}. More steeply evolving evolution may be expected in the context of BBH origin in star clusters that preferentially form early (globular clusters in particular, or nuclear star clusters), for which the formation of binaries may favor large delay times for early ejected binaries \citep{1993Natur.364..423S} and small delays for those retained \citep{2010MNRAS.402..371B}. More recent N-body simulations favor shorter BBH coalescence timescales for the most massive star clusters, $>10^6M_\odot$ due to the high escape velocity, so that BBH binaries are typically retained \citep{2013ApJ...763L..15M,2018PhRvD..98l3005R}, allowing the perturbing effects of stellar encounters in dense cores, leading to earlier BBH coalescence. This star cluster channel may deserve more exploration in the context of lensing, and with the growing interest in hierarchical BH growth for understanding the most massive BBH events \citep{2019MNRAS.487.2947D, 2020ApJ...896L..10R}.

\

While our paper was in preparation, a new study  \citep{Fishbach:2021mhp} has explored the delay time from individual GW events, including the released O3a data, which imposes a bound on the delay time parameter of $t_d^{eff}<4.5$ Gyr at $90\%$ CL, under the assumption that lensed events are not present, in agreement with recent results \citep{Abbott:2021iab}. Similar constraints $t_d^{eff}= 6.7_{-4.74}^{+4.22}$ Gyr on the delay time parameter are also obtained from the upper limit of the O3  stochastic GW background data \citep{Mukherjee:2021ags}. This limit and subsequent tighter constraints using the full GW data will translate into a lower bound on the lensing event rate for GW sources, given the assumed star formation rate evolution. Future studies that include both the unlensed and lensing detections will provide a self-consistent estimate of the delay time parameter with implications for the formation route of GW binary events, which are likely to have dependence also on the host environment \citep{2010ApJ...716..615O,2010MNRAS.402..371B, 2012ApJ...759...52D, Dominik:2014yma, Lamberts:2016txh, Elbert:2017sbr, Eldridge:2018nop, Buisson:2020hoq, Santoliquido:2020axb}.

%FINAL COMMENT

\section*{Acknowledgement}
 {The authors would like to thank the referee for very useful suggestions on the paper.} The authors are very thankful to Martin Hendry for carefully reviewing the manuscript and providing very useful comments which helped in improving the paper. S.M. also acknowledges fruitful discussion with Salvatore Vitale during a presentation in the LVK CBC telecon. S. M. is thankful to Benjamin D. Wandelt for his fruitful comments on the manuscript. 
This work is part of the Delta ITP consortium, a program of the Netherlands Organisation for Scientific Research (NWO) that is funded by the Dutch Ministry of Education, Culture, and Science (OCW). J.M.D. acknowledges the support of project PGC2018-101814-B-100 (MCIU/AEI/MINECO/FEDER, UE) Ministerio de Ciencia, Investigaci\'on y Universidades.  This project was funded by the Agencia Estatal de Investigaci\'on, Unidad de Excelencia Mar\'ia de Maeztu, ref. MDM-2017-0765.   This analysis was carried out at the Horizon cluster hosted by Institut d'Astrophysique de Paris. We thank Stephane Rouberol for smoothly running the Horizon cluster. We acknowledge the use of following packages in this work: Astropy \citep{2013A&A...558A..33A,2018AJ....156..123A}, IPython \citep{PER-GRA:2007}, Matplotlib \citep{Hunter:2007},  NumPy \citep{2011CSE....13b..22V}, and SciPy \citep{scipy}. The authors would like to thank the  LIGO/Virgo scientific collaboration for providing the noise curves. LIGO is funded by the U.S. National Science Foundation. Virgo is funded by the French Centre National de Recherche Scientifique (CNRS), the Italian Istituto Nazionale della Fisica Nucleare (INFN), and the Dutch Nikhef, with contributions by Polish and Hungarian institutes. This material is based upon work supported by NSF’s LIGO Laboratory which is a major facility fully funded by the National Science Foundation.

 \section*{Data Availability}
The data underlying this article will be shared at a reasonable request to the corresponding author. 
\vspace{-0.5cm}
\bibliographystyle{mnras}
\bibliography{main}
\label{lastpage}
\end{document}